# Evolutionary Design: Philosophy, Theory, and Application Tactics

V.V. Kryssanov[1], H. Tamaki[2], S. Kitamura[2]

[1] College of Information Science and Engineering, Ritsumeikan University, Kusatsu, Japan
[2] Faculty of Engineering, Kobe University, Kobe, Japan

**Abstract**
Although it has contributed to remarkable improvements in some specific areas, attempts to develop a universal design theory are generally characterized by failure. This paper sketches arguments for a new approach to engineering design based on Semiotics – the science about signs. The approach is to combine different design theories over all the product life cycle stages into one coherent and traceable framework. Besides, it is to bring together the designer's and user's understandings of the notion of 'good product'. Building on the insight from natural sciences that complex systems always exhibit a self-organizing meaning-influential hierarchical dynamics, objective laws controlling product development are found through an examination of design as a semiosis process. These laws are then applied to support evolutionary design of products. An experiment validating some of the theoretical findings is outlined, and concluding remarks are given.

## 1 INTRODUCTION

Historically, design tasks have been verifying theories, challenging their value in a number of domains varied from quite 'objectivistic' but highly abstract, such as philosophy and mathematics, to rather situated but pragmatic, such as economics and management, and further to essentially subjective and loosely structured, such as art and sociology. It is then not surprising that in the present era of technocratic civilization, design receives significant research attention, having been studied first as a craft and later – as a field of engineering that promises to eventually grow into an 'independent' discipline developed in line with the traditions of classical science. In an effort to contribute to the development of this discipline, both academic researchers and practitioners, although seeing the problem from fairly different perspectives, have been trying to understand and justify if and how one or another theory can be useful for solving design tasks.

Adherents of mathematical approaches habitually assess the usefulness of a theory from the standpoint of its logical consistency and tractability: only those theories are considered good, which allow for the generation of true theorems. When it comes to product design, such theorems can be reduced to statements of the following form: 'a design is right if $\pi$,' where $\pi$ is a formally determinable quality or property (see, for example, reference [1] and, to some extent, reference [2]). The role of the theories is seen to build an apparatus (terminology, axiomatic basis, inference methods, etc.) for the design science, while their main target is to achieve better control over the design process. However, there are many obstacles preventing the broad introduction of these theories into practice. The most fundamental obstacle is the ontological contradiction between the notions of 'objectively' good (i.e. right) design and always subjectively good product. This contradiction makes the development of a 'general' design theory hardly feasible unless a universal notion of 'good' product is formally clarified.

Proponents of a different viewpoint claim that theories are good if they allow for understanding design processes and products better. Having an obvious analytical flavor, these theories are expected to incorporate a large amount of domain-specific knowledge to assist the designer in his or her decision-making (e.g. [3, 4]). It is assumed that the quality of a product solely depends on knowledge possessed by the designer, and the problem of designing good products is, in the main, a problem of professional expertise. A weak point here is that possessing even 'ideal' knowledge is neither necessary nor sufficient condition for the creation of a commercially successful product, while the history witnesses that individuals with little qualification and experience but well developed creative abilities invented many great products.

In the spirit of recent advances in psychology, ethnomethodology, anthropology, and cognitive sciences, some put forward suggestions, highlighting a need for observationally and analytically studying gifted people, who seem to be able to design, devise, and develop without any systematic guidance from the outside (see reference [5]). It is believed that research of this kind would ultimately lead one to discovering a set of thinking techniques and methodologies that can be taught to designers to make them designing better products (e.g. [6]). An obvious danger, however, is that apart from the usual difficulties with the application of any cognitive model, many factors affecting designing will necessarily be missed in the resultant techniques because of the complexity and instability of the 'network' topology of social and psychological relationships: the process of product concept creation is not limited by activities at the designer's side, and it is generally not homogeneous in terms of space- as well as time-situated interactions [7]. The latter could question the applicability and usefulness of the obtained techniques at all.

It is now well recognized that there are multiple perspectives on the building of the design discipline, each of which contains part of the truth, but none of which contains the entire solution [8]. Leaving alone marginal and unconstructive claims like 'to design, we should only learn from nature' or 'design is art and never be science,' which nonetheless reflect opinions of some professionals, a principal question then arises: whether it is possible to bridge all the different viewpoints and attitudes toward design within one scientific framework, and if so – what would be the basis for that?

It is our strong belief that Semiotics – a science about signs, sign systems, and signification – is exceptionally well suited to link the mathematical, social, and cognitive theories together to provide insights about design as a single discipline and encourage the design community to develop new methods and tools to aid designers and lighten the routine burden but improve the quality and 'goodness' of products. To justify this supposition, we will discuss a semiotic approach to the design study and illustrate it with an example in the following sections. This paper continues the investigation, which the authors started with the Semiotic Theory of Creativity for engineering design [9] that later gave rise to the development of a more general Semiotic Theory of Evolutionary Design [10].

## 2 SEMIOTICS AND THE 3-LEVEL PARADIGM

Perhaps the most significant thing missing from the traditional design theories is the failure to appreciate how products are conceived by people as opposed to simply perceived by people. Indeed, a product, as we comprehend it in everyday life, is always more than just an artifact – a physical entity to which human must adapt, while all the product life-cycle processes have a complex, meaning-influential dynamics with regard to the product, its concept, and use-environments. People, through their activities and practice, originate an 'objective' need for the creation of the product by subjectively assigning its **intended** meaning and, at a later time but also subjectively, evaluate the product's value in an environment, i.e. its **actual** or **emergent** meaning, yet having many different views and opinions while doing so. Semiotics studies this essentially **meaning-making** process, where the product among other phenomena of the environment is construed as a sign, which needs to be interpreted to allow for the use of the product throughout its life cycle.

By Peirce, Semiotics deals with three subjects [11]: the **representamen** – the sign itself, the **object** – that which is signified, and the **interpretant** – the meaning that follows semantically from the process of interpretation of the representamen. The main postulate of Peircean Semiotics is that no representamen directly points us to an object, and a sign has a meaning only for a system of interpretance that is, by its very nature, a system of other signs – a **sign system**. Therefore, a representamen is a sign of an object only for some sign system and not necessarily for all sign systems: the same representamen may signify a different object, or the same object may be signified by a different representamen, etc. – the variety of possible combinations is not limited by one-to-one relations.

Modern biology, sociology, and physics teach us that the richness of the complexity of many natural systems derives from a strategy to organize smaller units into larger ones, which are in turn to be arranged into still larger ones, and so on [12]. As we shall further see, all the meaning-making processes – the **semiosis** processes – reveal a similar hierarchical organization, and sing systems generally have a multi-level structure subject to these processes.

Let us assume that in a multi-level sign system, signs on level N are dynamically composed of signs of level N-1 so that only those of all the possible combinations of that lower-level signs occur, which are allowed by boundary conditions set at level N+1; signs at level N-1 are then **constitutive** for level N signs, and signs at level N+1 are **constraining** for level N. The dynamics of semiosis can be described in terms of the interactions among the adjacent levels of such a sign system [13].

Most naturally for the designer, semiosis is a material process, which necessarily involves a translation or re-representation – re-interpretation – of information from one meaning-making level to another. Signs on level N can be representamina of the objects and phenomena at level N-1 for processes and structures at level N+1, which form a system of interpretance. The lowest-level signs (e.g. physical objects, phenomena, behavioral dispositions, emotions, and the like) are perceived or realized through their **distinctions** and get a representation at the 'intermediary' level of the designer's cognition in respect to the interpretive laws of the highest, experiential and environmentally (culturally, socially, technically, economically, etc.) induced level, which accommodates interpretants and assigns a meaning to the representamina. The design process is the process of introducing a sign at level N with a meaning for both level N-1 (as an entity materially grounded there) and level N+1 (as an entity having a material relevance at this level). Generally, the sign is allowed to have many possible meanings, depending on contextual constraints from higher interpretive levels.

People perceive a completed product through its distinctions that need to be interpreted and may acquire a meaning for a sign system. Although human perception is relatively uniform and consistent, it is natural that the meaning assigned to a representamen can vary significantly, depending on the subjective dynamics of perception (level N) and interpretation (level N+1) as well as on relations between the adjacent semiotic levels.

With a potentially infinite hierarchy of interpretative levels, where signs on level N+1 can in their turn be constitutive, i.e. be objects for a higher level and so on, a product is perceived and conceived in the same (ensemble of) sign system(s) that makes up a **language** (here understood in a very general sense and not limited to handling verbal constructions). A product can be considered as a 'text' written in this language, which has a **syntax** constraining the product organization, **semantics** defining the meaning(s) of the product, and **pragmatics** reflecting various use-effects (e.g. physiological, psychological, or social) associated with the product. The design science may be seen as a science about the evolution of the language – a science that studies fundamental laws of the semiosis processes, which govern the product life cycle.

## 3 ELEMENTS OF THE THEORY

### 3.1 Industrial Semiosis

The concept of Industrial Semiosis categorizes the product life-cycle processes along three semiotic levels of meaning emergence [14]: 1) the **ontogenic** level that deals with the life history data and future expectations about a single occurrence of a product; 2) the **typogenic** level that holds the processes related to a product type or generation; and 3) the **phylogenic** level that embraces the meaning-affecting processes common to all of the past and current types and occurrences of a product. The three levels naturally differ by the characteristic durational times of the grouped semiosis processes: as one moves from the lowest, ontogenic level to the higher levels, the objects become larger and more complicated and have slower dynamics in both original interpretation and meaning change.

A product (as concept) starts its development with initially coinciding onto-, typo-, and phylogenesis processes but distinct and **pre-existing** semiotic levels of interpretation. The concept is evolved, and typogenesis works to re-organize the relationships between the onto- and phylogenesis processes, as the variety of objects involved in product development increases. Product types and their interactions mediate – filter and buffer – between the levels above and below: not all variety of distinctions remains available for re-organization as *phylos*, nor every lowest-level object have a material relevance there. The phylogenic level is buffered against variations at the ontogenic level by the stabilizing mediations at the typogenic level. (Note that all the three levels of product definition are to mediate between more global levels of product environments.)

The dynamics of the interactions between the semiotic levels can well be described in terms of the basic processes of **variation** and **selection** [15]. In complex system evolution, variation stands for the generation of a variety of simultaneously present, distinct entities (**synchronic variety**), or of subsequent, distinct states of the same entity (**diachronic variety**). Variation makes variety increase and produces more distinctions. Selection means, in essence, the elimination of certain distinct entities and/or states, and it reduces the number of remaining entities and/or states.

From a semiotic point of view, the variety of a product intended to operate in an environment is determined by the devised product structure (i.e. the relations established between product parts – its synchronic variety) and the possible relations between the product and the anticipated environment (i.e. the product feasible states – its potential diachronic variety), which together aggregate the product possible configurations. The variety is defined on the ontogenic level that includes elements for description of both the structure and environment. The ontogenesis is driven by variation that goes through different configurations of the product and eventually discovers (by distinction selection at every stage of the product life cycle) configurations, which are **stable** on one or another time-scale. A constraint on the configurations is then imposed, resulting in the selective retention – emergence of a new meaning for a (not necessarily new) sign – at the typogenic level. The latter decreases the variety but specializes the ontogenic level so that only those distinctions ultimately remain, which fit to the environment (i.e. only dynamically stable relation patterns are preserved). Analogously but at a slower time-scale, the typogenesis results in the emergence of a new meaning on the phylogenic level that consecutively specializes the lower levels. Thus, the main semiotic principle of product development is such that the dynamics of the meaning-making processes always seeks to decrease the number of possible relations between the product and its environment and hence, the semiosis of product life cycle is naturally simplified. At the same time, however, the 'natural' dynamics is such that augments the evolutive potential of the product concept for increasing its organizational richness: the emergence of new signs (that may lead to the emergence of new levels of interpretation) requires a new kind of information and new descriptive categories must be given to deal with the still same product.

### 3.2 On Formalization

Among many possible approaches to formalization of design semiosis, we have chosen Algebraic Semiotics (see reference [16]) for its expressiveness, clarity, and freshness. Algebraic Semiotics deals with signs as members of sign systems defined to be algebraic theories with extra structure, and semiosis processes are specified through semiotic morphisms to be a kind of mapping of algebraic theories. Instead of defining properties of a sign system by reference to its members, Algebraic Semiotics, unlike other approaches but Category Theory, does so by reference to its external relationships with other sign systems.

A sign system is represented as a five-tuple $\Xi = \langle S, V, C, R, A \rangle$, where $S$ is a sort-set for signs in the system, $V$ is a sort-set for data, $C$ is a set of operations called 'constructors' that are used to create signs from other signs, $R$ is a set of relations defined on the system signs, and $A$ is a set of axioms that constrain the possible signs. $S$ and $C$ are partially ordered: by subsort and by level, respectively; in turn, constructors are partially ordered by priority within each level.

A semiotic morphism $M: \Xi \rightarrow \Xi'$ is a translation that consists of partial functions, which map sorts, constructors, predicates and functions of a sign system $\Xi$ to sorts, constructors, predicates and functions of a sign system $\Xi'$ and retain some of the structure of $\Xi$: the mapping of sorts to sorts preserves arguments and result sorts of constructors and predicates as well as the subsort ordering, and it does not change data sorts.

### 3.3 Semiosis Laws

In this section, we will put forward a bold conjecture that (the dynamics of) all the life-cycle meaning-making processes can be described in terms of **basic semiotic components** – algebraic constructions of the following form:

$$P_n(M_n: f_n [\Xi_n] \rightarrow \Xi_{n+1}), \tag{1}$$

where $\Xi_n$ is a sign system corresponding to a representation of a (design) problem at time $t_1$, $\Xi_{n+1}$ is a sign system corresponding to a representation of the problem at time $t_2$, $t_2 > t_1$, $f_n$ is a composition of semiotic morphisms that specifies the interaction of variation and selection under the condition of information closure, which requires no external elements be added to the current sign system; $M_n$ is a semiotic morphism, and $P_n$ is the probability associated with $M_n$, $\Sigma P_n = 1$, $n=1,\ldots,M$, where M is the number of the meaningful transformations of the resultant sign system after $f_n$. There is a partial ranking – importance ordering – on the constraints of $A$ in every $\Xi_n$, such that lower ranked constraints can be violated in order for higher ranked constraints to be satisfied. The morphisms of $f_n$ preserve the ranking.

The Semiotic Theory of Self-Organizing Systems postulates that **in the scale hierarchy of dynamical organization, a new level emerges if and only if a new level in the hierarchy of semiotic interpretance emerges** [13]. As the development of a new product always and naturally causes the emergence of a new meaning, the above-cited Principle of Emergence directly leads us to the formulation of the first law of life-cycle semiosis as follows:

I. The semiosis of a product life cycle is represented by a sequence of basic semiotic components, such that at least one of the components is well defined in the sense that not all of its morphisms of $M$ and $f$ are isomorphisms, and at least one $M$ in the sequence is not level-preserving in the sense that it does not preserve the original partial ordering on levels.

For the **present** (i.e. for an on-going process), there exists a probability distribution over the possible $M_n$ for every component in the sequence. For the **past** (i.e. retrospectively), each of the distributions collapses to a single mapping with $P_n=1$, while the sequence of basic semiotic components is degenerated to a sequence of functions. For the **future**, the life-cycle meaning-making process can be considered in a very general probabilistic sense only (e.g. in terms of probability distributions that are characteristic of a specific domain, social group, design approach, or the like).

It seems logical to assume that the successful (perhaps, in any sense) introduction of a product to the market effects the introduction and settlement of the corresponding meanings at the onto-, typo-, and phylogenic semiotic levels. Let us denote the number of relations between the product and its environment as ε. We can now formulate the second law of life-cycle semiosis as follows:

II. A component $P_n(M_n: f_n [\Xi_n] \rightarrow \Xi_{n+1})$ represents a **successful** life-cycle semiosis process if the morphism $M_n$ is **natural** in the sense that $\varepsilon_n > \varepsilon_{n+1}$.

Although the above laws have been formulated with sufficient precision, it is recommended to apply them (alike Algebraic Semiotics in general) in an informal way, calling for details only in boundary and difficult situations. The main purpose of these as well as other not-yet-formulated laws of life-cycle semiosis is to guide the examination of the product development and usage processes, no matter which design theory or even paradigm is employed at the lower, applied level.

## 4 EVOLUTIONARY DESIGN

Evolutionary Design is a relatively new paradigm that encompasses the recently popular design approaches, such as Sustainable Design, Design for X, Green Design, and the like, which explicitly recognize the social, evolutionary, and error-prone nature of product development and postulate the tentative character of design solutions, making them dependent on the dynamics of product use-environments. In the following paragraphs, we will give a semiotic interpretation for this paradigm and show how the laws of life-cycle semiosis could be applied to support evolutionary design.

### 4.1 Semiosis of Evolutionary Design

Every design is based on some expectations, which explicitly or implicitly determine the product intended meaning and which are realized as design requirements – once conceived relation patterns. These expectations always fit particular environmental conditions, which often become obsolete before the product reaches into the market place. A universal and obvious solution to this problem is to increase as much as possible the synchronic variety of the product by contriving appropriate decisions in design (the notorious idea of re-configurable products). Indeed, the more elaborate the structure of the product (or its concept), the larger the number of environmental situations in which it can maintain. Different product configurations can fit (or be adapted to) different situations and, therefore, in the case of dynamic environments, design evolution should increase the synchronic variety, making the product more complex to adequately react to the environmental changes.

Although the latter statement does not contradict the life-cycle semiosis laws and is, perhaps, true in general, this does not mean that the 'best' product must always be the most complex one, i.e. be the product with the maximal synchronic variety. Due to many reasons – economical (costs), technical (reliability), ecological (energy and material consumption, pollution), social and ergonomic (safety, convenience and easiness in production and operation), etc., the best is the product with the simplest possible structure for the given functionality, i.e. with the least possible (for the given environment) synchronic variety. In this sense, the 'goodness' or, better say, adequacy of the product depends on the characteristics of product environment in relation to the implemented design expectations, i.e. it depends on how well the intended meaning matches the meanings emerging through onto-, typo-, and phylogenesis (if any).

Design expectations can roughly be classified into two categories [17]: *a*) functional – the expectations about operation of the product and its functional parameters, and *b*) environmental – the expectations about the product-environment interaction. In the product life cycle, the distinction dynamics is driven by violations of design expectations. The dynamics of ontogenesis processes, where relation patterns are originally detected to be further interpreted and accepted or rejected for an action, is subject to psychological and physiological laws, such as, for instance, the well-known law of Weber-Fechner. Having been differentiated by the time-scale of the corresponding meaning-making processes, violations of functional expectations control the product typogenesis, while violations of environmental expectations influence both the typo- and phylogenesis semiosis processes.

Resolving the product intended meaning mismatch is a critical task in design and life cycle engineering that requires the development of the appropriate information technologies and tools. Below, we will outline an agent-based technology that was developed to detect violations of design expectations and support, in this way, evolutionary design of high-tech products, as well as to assess the successfulness of the life cycle of an individual product, product type, or product family on the whole (also see reference [18]).

### 4.2 Computer-aided Evolutionary Design

The main idea of the developed technology of evolutionary design support is to allow for evaluation and change of the most important design expectations using programmable mobile agents, called expectation agents, which utilize design requirements represented explicitly to monitor product functionality, usage, and operational environment. An expectation agent typically consists of a static hardware unit (including transducers, processing and preprocessing blocks, etc.) integrated with the product and a mobile software part, which allows the agent to demonstrate a certain level of autonomy, intelligence, and proactivity in respect to the product and act on behalf of the designer or manufacturer. The agent can execute various control procedures, transfer the registered data via a communication line, assist the product user, collect feedback directly from the user, and update its own code. Data obtained with the agent can be analyzed and used to detect when the usage and environmental patterns shift thereby necessitating optimization of the synchronic variety of the product by adjusting its configuration. To provide for the evolutionary support of the life cycle semiosis processes, an effective information infrastructure connecting products supplied with agents, service centers, manufacturers, and designers is to be developed. Figure 1 gives an example of the agent networking that would be arranged using, for instance, the existing infrastructure of the World-Wide Web.

There are three distinct but overlapping layers of networking driven by the onto-, typo-, and phylogenesis meaning-making processes. At the ontogenic layer, expectation agents monitor individual products and their actual environments. The agents communicate with each other as well as with the design, manufacturing, maintenance, and other involved parties. They create product life histories and try for optimization of technical, technologic, and usage processes associated with the products. Both the product operation and the expectation agent functionality can be changed by updating the agent's program code. The ontogenic layer comprises single products installed in dynamic environments. An essential feature of the layer is that no order is imposed from the outside on the agents' communication, but the agents are to be evolutionary grouped by the environmental characteristics into specific clusters. The emergence of a cluster at the typogenic layer implies the emergence of a new meaning – a product type – at the corresponding level of semiotic interpretance. In an ideal case, the number of active clusters indicates the number of product types requisite for the given environment covered by the agent network.

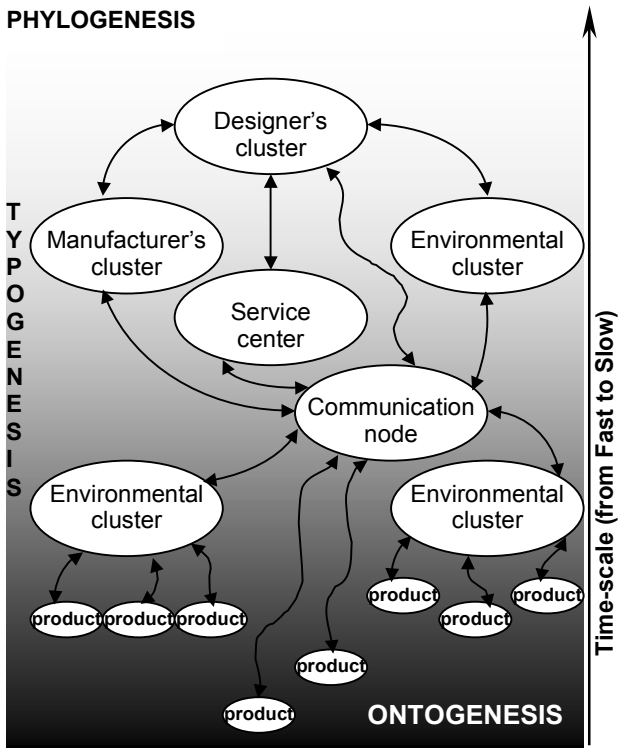

Figure 1: Expectation agent networking.

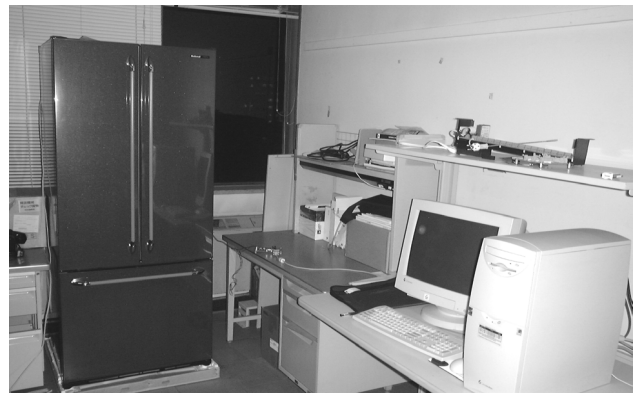

Figure 2: Refrigerator with an installed expectation agent.

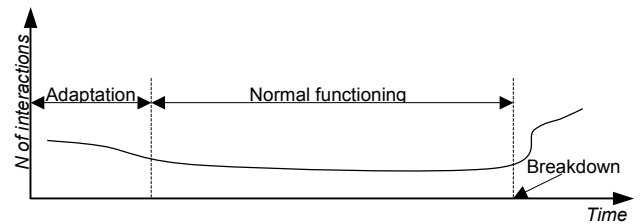

Figure 3: The dynamics of product-environment interaction.

The typogenic layer provides for systematization of the design information flows generated by the agents. Information and data are sorted and stored, depending on the product type (version or generation). A majority of actions and processes of this layer are defined on a population of products grouped into an environmental cluster. Of course, environmental clusters may naturally be grouped into still larger units of characteristic dynamics, e.g. by different manufacturers. Such upper-scale grouping would initiate the emergence of a new meaning – product family – at the phylogenic level of interpretance.

The phylogenic layer links all the design, production, maintenance, utilization, etc. processes required for or associated with the realization of a distinct product or technology concept. Information flows of this layer are relatively stable and depend on the global (e.g. cultural and ecological) rather than specific (e.g. technical and economic) factors. The first law of life-cycle semiosis regulates the main processes of phylogenesis and makes the agent networking really possible.

Figure 2 presents a product – refrigerator – with an embedded expectation agent that is part of the experimental setup built through our study. The setup is to compare the behavior of virtual products (i.e. product intended meanings reconstructed with virtual objects) and real products by means of multi-media, haptic devices, and remote sensors (see reference [10] that gives a detailed account of the experiment). An analysis of empirical data collected with the setup showed us that the dynamics of product-environment interactions could serve as an indicator of the 'successful' (for the given environment) operation of the product.

Figure 3 depicts the characteristic change of the number of product-environment interactions, such as product part movements and working mode switchings, registered by the expectation agent (high-frequency fluctuations have been filtered out of the data set by a pre-processing block of the agent). Although it is admitted that a larger scale empirical study is required to prove the efficiency of the evolutionary design support based on the expectation agent networking, the obtained results principally confirmed our theoretical conjectures and demonstrated the technological feasibility of the approach (also, see reference [17] that discusses the case of violations of design expectations as well as possible strategies to react to detected violations in the light of the Evolutionary Design Theory).

## 5 CONCLUSIONS

We would like to conclude this paper by the following remarks.

First, the idea of semiotic interpretation of the design process is not new; moreover, it has become almost a 'fashion' for the design community in the last few years, and there are many publications on this subject. One principal difference of our work is that we have not limited the investigation by a semiotic analysis and classification of design objects and signs representing them but, instead, focused on the processes responsible for the development of these objects and their meanings.

Second, we see the rôle of the semiotic approach to design as not to discriminate true from false, correct from incorrect, bad from good, etc., but to provide designers with a new perspective on design theories and techniques, to better understand how this or that process goes on, and what are the factors, subjective and objective, affecting it. Further elaboration of the semiotic approach will, we believe, shed light upon many non-obvious consequences and causes of the application of a particular design theory or technique.

Finally, we come clean that Semiotics alone cannot account for all the results and insights brought out by all the design and life cycle theories. Rather, Semiotics can help us merge all the stages of product development together within a uniform and universal scientific framework. The latter could be seen as the ultimate goal of our research.

## 6 ACKNOWLEDGMENTS

The authors would like to acknowledge the financial support of the Japan Society for the Promotion of Science (project No 96P00702) that made this research possible. We are greatly indebted to Dr. I. Goncharenko for his key contribution of the experimental part of the study.